\documentclass[sigconf, screen]{acmart} 

\setcopyright{none}
\acmDOI{}
\acmISBN{}

\acmConference[LIMITS '26]{12th Workshop on Computing within Limits}{June 23--25, 2026}{Online}

\usepackage{times,latexsym}
\usepackage{xurl}
\usepackage{graphicx}
\usepackage{subcaption}
\usepackage{multirow}
\usepackage{makecell}
\usepackage{booktabs}

\usepackage{todonotes}

\newcommand{\KILL}[1]{}

\usepackage{glossaries}
\newacronym{nlp}{NLP}{Natural Language Processing}
\newacronym{llm}{LLMs}{Large Language Models}
\newacronym{ml}{ML}{Machine Learning}
\newacronym{DL}{DL}{Deep Learning}
\newacronym{ICT}{ICT}{Information and Communication Technology}
\newacronym{PUE}{PUE}{Power Usage Efficiency}
\newacronym{TDP}{TDP}{Termal Design Power}
\newacronym{AI}{AI}{Artificial Intelligence}
\newacronym{LCA}{LCA}{Life Cycle Assessment}
\newacronym{IC}{IC}{Integrated Circuits}

\newcommand{\COtwo}{CO\textsubscript{2}~eq}
\newcommand{\Sbe}{Sb~eq}

\begin{document}

\title{The Rising Unsustainability of AI Graphics Cards Production}

\author{Cl\'ement Morand}
\email{clement.morand@lisn.fr}
\orcid{0009-0007-8575-9874}
\affiliation{%
    \institution{Université Paris-Saclay, CNRS, LISN}
    \city{Orsay}
    \country{France}
}

\author{Aur\'elie N\'ev\'eol}
\email{aurelie.neveol@lisn.fr}
\affiliation{%
    \institution{Université Paris-Saclay, CNRS, LISN}
    \city{Orsay}
    \country{France}
}

\author{Anne-Laure Ligozat}
\email{anne-laure.ligozat@lisn.fr}
\affiliation{%
    \institution{Université Paris-Saclay, CNRS, LISN, ensIIE}
    \city{Orsay}
    \country{France}
}

\date{\today}

\begin{abstract}
The rapid advancement of Artificial Intelligence (AI) has been accompanied by significant increases in computational and environmental costs, driven by large-scale investments in AI infrastructure, hardware, and software. In particular, graphics cards have become central to AI training, with frequent hardware updates required to meet escalating computational demands. However, the environmental damages of graphics cards production remain understudied.

This study addresses this gap by estimating the environmental damages associated with graphics cards production over the past decade (2013–2025). We analyze trends in energy consumption, carbon emissions and resource depletion.

We compile and provide a dataset documenting the environmental damages of NVIDIA workstation graphics cards production since 2013. Our analysis of this dataset reveals a steady increase in production-related impacts over the period. 

Our finding highlights the need for greater transparency in life-cycle data, a persistent challenge in AI environmental assessments. While operational efficiency improvements (e.g., energy-efficient training, carbon-aware computing) are often prioritized, our results underscore that production-related impacts are also escalating and cannot be overlooked. 

The AI community must move beyond incremental optimizations and confront the necessity of sufficiency. This shift may demand structural changes such as policy interventions, hardware design for longevity, and cultural shifts away from perpetual growth and increased performance.

\end{abstract}

\keywords{AI, graphics cards, embodied impacts, carbon footprint, resource depletion, environmental trends, Life Cycle Assessment, computational limits, environmental damages, sustainability}

\maketitle

\section{Introduction}

The growing carbon footprint of \gls{ICT} is documented~\citep{Freitag2021real,Malmodin2023ICT}, with a relatively even distribution of greenhouse gas emissions related to user devices, networks, and datacenters. In recent years, a large hype around \gls{AI} with generative general purpose large language models has unfolded~\citep{Varoquaux2025bigger}. This has lead to massive investments and deployments of AI infrastructures, equipment and software, causing concerns around the unsustainability of such deployment in a sector already following an unsustainable trajectory. Campaigns like Save the AI \url{https://savethe.ai/} or reports like the one from the Shift project~\citep{Shift2025IA} alert on the energy and water consumption of AI. Big tech companies have presented AI as one of the main drivers of the growth of their carbon footprint and preventing them from attaining their (now dismissed\footnote{See for instance \url{https://www.tomshardware.com/tech-industry/google-quietly-removes-net-zero-carbon-goal-from-website-amid-rapid-power-hungry-ai-data-center-buildout-industry-first-sustainability-pledge-moved-to-background-amidst-ai-energy-crisis} and \url{https://26227852.fs1.hubspotusercontent-eu1.net/hubfs/26227852/Why\%20Companies\%20are\%20Dropping\%20Net-Zero.pdf} [Last accessed 02/04/2026].}) decarbonation targets~\citep{Google2024environmentalreport,Meta2024environmentalreport}.

Different studies have documented evolution in datacenter energy consumption, and assessed the role the current AI hype plays~\citep{Masanet2020recalibrating,Shehabi2024USDatacenterreport,Shift2025IA,IEA2024electricity}. They show increases in demand that are not compensated by increases in datacenter efficiency anymore and highlight the preponderant share of the increases in energy consumption that are related to AI activities. 

Datacenters use a large quantity of servers for storage and compute, in addition to networking equipment, cooling infrastructures and electrical installations. 
Their environmental impacts thus do not only occur during usage but also during the complete life cycle of these appliances. In particular, if servers environmental damages are relatively well understood, notably thanks to parametric models like boaviztapi~\citep{Simon2024boaviztapi}, graphics cards and other AI accelerators production damages remain little documented, with almost no quality data existing until the last two years. This lack of information means that we cannot fully grasp the unsustainability of the AI sector and its material implications, especially as large datacenter projects have been announced in a lot of global north countries with hundred thousands of graphics card per installation\footnote{See for instance the data on "frontier data centres" by Epoch AI at :\url{https://epoch.ai/data/data-centers} [Last accessed 02/04/2026] or the cartography of datacenters and their contestations in France by \textit{Le nuage était sous nos pieds} at: \url{https://lenuageetaitsousnospieds.org/articles/2025-12-11-carte-des-data-centers-des-projets-et-des-contestations-en-france.html} [Last accessed 02/04/2026].}. 


AI hype, and the massive growth in ICT infrastructure it brings is at odds global sustainability issues. Big tech companies dismissing their decarbonation targets is in direct contradiction with the pressing need to align human activities with global limits, as exemplified with the notion of planetary boundaries \citep{Steffen2015planetary}. 
\citet{Nardi2018LIMITS} proposes ways forward for research on computing within limits. They insist on the need to question growth, consider models of scarcity, and work toward reducing energy and material consumption. 
Better understanding the environmental implications of graphics cards, and their evolution in time and in procurement, thus seems very important to better understand the unsustainability of AI, questioning its massive growth. Furthermore, a lot of impact reduction strategies are focused on greening electricity supplies of ICT infrastructures. These strategies do not allow to reduce production impacts, which may occupy a growing share of the environmental damages of these infrastructures. 

Specifically, we investigate how the production damages of individual graphics cards have evolved in the past 12 years.
We scrutinize the characteristics and environmental impacts of the production of NVIDIA workstation graphics cards over (2013-February 2025).
In summary, our work makes the following primary contributions:\footnote{The Dataset on graphics cards production damages, and the code used in the experiments, and all necessary information for reproducibility are available at: Morand, C. (2026). \textit{Dataset on NVIDIA workstation graphics cards from 2013 to 2024 and their estimated environmental impact} (Version v2) [Data set]. 12th Workshop on Computing within Limits (LIMITS'26), Online. Zenodo. \url{https://doi.org/10.5281/zenodo.20321644}.} 

\begin{itemize} 
    \item 
     A dataset documenting the environmental damages of producing graphic cards
     released between 2013 and 2025; 
    \item 
    Evidence that the impact of graphics cards has increased consistently over the period; 
    \item 
    Trends in the footprint of graphics cards procurement.
\end{itemize}

\section{Related Work}

Corporate reports of the environmental damages of \gls{ICT} equipment often rely on \textit{Product Carbon Footprint} (PCF) sheets, using an assessment methodology that lacks transparency. \gls{LCA} is considered more robust is increasingly used by researchers and industry players. Industry players have for instance used \gls{LCA} to report on the impact of servers produced by Dell~\citep{sphera2021lca,thinkstep2019lca}. Google also recently used this methodology to release an assessment of the carbon foorpting of the TPUs they produce~\citep{Google2025acvTPU}.

\gls{LCA} has been used by researchers, especially to document the environmental damages associated with the different components of \gls{ICT} equipment.
User and \textit{Internet of Things} (IoT) devices~\citep[e.g.,][]{Gupta2020chasing,Clement2020sources,Pirson2023Evaluating}, servers~\citep[e.g.,][]{Loubet2023life,Simon2024boaviztapi,Groger2021green} or more specific components like \gls{IC}~\citep[e.g.,][]{Roussilhe2024silicon,Pirson2022environmental,Pirson2023environmental} have thus been documented. These studies have shown that production represents a significant share of the carbon footprint of servers~\citep{Morand2024MLCA,Simon2024boaviztapi} and the vast majority of carbon footprint for user devices~\citep{Gupta2020chasing} over their life-cycles. They have highlighted the prevalent role of \gls{IC} at the component level~\citep{Clement2020sources} and their intensifying production damages for newer generations~\citep{Pirson2022environmental,Pirson2023environmental}. They have also shown that damages are not limited to greenhouse gas emissions, with for instance important metallic resource depletion concentrated on the production phases of hardware~\citep{Morand2024MLCA,falk2025carboncradletograveenvironmentalimpacts}. Finally, they have also shown that data sources have significant impact on the obtained assessments~\citep{Pirson2022environmental,wattiez2024exploring} and that keeping up to date geographically, temporally and technologically representative data for \gls{ICT} is very complex notably because of the rapid evolution of the sector and the lack of transparency by manufacturers~\citep{Clement2020sources}.

This has lead to initiatives aimed at developing open source, reliable and open access commons on the environmental damages of ICT, such as within the Boavizta association\footnote{\url{https://boavizta.org/} [Last accessed 02/04/2026]}, as well as parametric models for easier assessment of the damages of a variety of devices with a fair precision~\citep{Groger2021green,Simon2024boaviztapi,Morand2024MLCA}.

Still, there is little information on graphics cards production impacts. \citet{Loubet2023life} performed an LCA of servers including a graphics card, but not of the type used for AI. \citet{Luccioni2022estimating} used a probable damage value for an AI graphics card, but did not conduct any study. \citep{Morand2024MLCA} and \citep{Berthelot2024generative} upgraded previous parametric models to also include graphics cards but not with good quality data especially on components other than the GPU and memory. Manufacturers have very recently started disclosing carbon footprint information, with Google releasing a life cycle assessment of some of their TPU chips~\citep{Google2025acvTPU}, and NVIDIA providing PCF sheets on their latests models~\citep{NVIDIA2025pcfH100,NVIDIA2025pcfB200}. Independent analyses have been conducted on the A100 SXM4 40GB card~\citep{falk2025carboncradletograveenvironmentalimpacts} which~\citet{ADEME2026GPU} also extends to a selection of other cards, either with a full analysis or a simplified one using a parametric model based on a few key characteristics.  
These new results are very welcome, but they remain limited to a relatively small number of cards, and do not necessarily allow to understand trends in the environmental implications of graphics cards production.

Few sources exists on global graphics card procurement. \citet{devries2023growing} discusses increases in AI energy consumption related to the surge in deployment and use of generative large language models, providing some figures on the expected sales of graphics card by NVIDIA over the next years. \citet{wang2024waste} attempts to model the evolution of hardware requirements associated with large language model deployment in order to assess the volume of e-waste it implies.  
\citet{Shift2025IA} provides estimates on the number of installed cards since 2019, based on assessments of the global installed compute power divided by the peak compute power of the dominant card every year. This data over a longer period is interesting but presents discontinuities when switching from one card generation to another, resulting in over-estimations of the number of cards in threshold years. Finally, \citet{EpochAIChipSales2026} provides disagregated information on the number of cards sales since 2022, based on financial data from big tech companies. This data source appears most reliable, but does not cover the largest time period.

Overall, existing studies 
accurately document the environmental damages of ICT equipment, even if data challenges remain. Parametric models have proven useful to document damages of a wide variety of equipment at a correct precision. Information on graphics cards in particular remain scarce. 
In this paper, we adapt an existing parametric model for graphics cards production damages~\citep{Morand2024MLCA} using recent information on graphics cards and combine it with a newly curated dataset on the characteristics of NVIDIA workstation graphics cards over the last decade to document the evolution in graphics cards production damages over the period. We then combine this data with information on the graphics cards sales to provide insights on the damages incurred by graphics card production at a global scale. 

\section{Material and Methods}
\label{sec:methods}

\subsection{Gathering Information on Graphics Cards}

\label{subsec:graphics-cards}

\begin{table*}[ht]
    \centering
    \begin{tabular}{cccccc}\toprule
                Minimum   &   First quartile  &     Median    &     Mean  &    Third quartile    &     Maximum \\\midrule
2013-01-05 & 2015-06-29 & 2018-03-27 & 2018-05-15 & 2021-04-30 & 2024-11-18\\\bottomrule
    \end{tabular}
    \caption{Distribution of models release date of NVIDIA workstation graphics cards (2013-2025)}
    \label{tab:distribution_workstation}
\end{table*}

\gls{LCA} of ICT equipment have shown the importance of \gls{IC} in the environmental impacts of ICT equipment \citep{Clement2020sources}. \gls{IC} come in two forms in graphics cards: GPU (logic type \gls{IC}) and memory (memory type \gls{IC}). The surface of the GPU is indicated by the GPU \textit{die area}. 
Contributors to the impact of producing \gls{IC} include the surface of the \gls{IC} (i.e., the die area of the GPU and the surface of memory type \gls{IC} for the memory chips), as well as how finely the circuits are printed on the semiconducting material. 
Thus, if the die area increases we can expect an increase in the environmental impacts of the device. Furthermore, \citet{Pirson2023environmental} have shown that, with finer technological nodes, the environmental impacts per produced cm$^2$ of die increase. Thus, as the quantity of memory increases (probable increase in memory type \gls{IC} surface), the GPU die size increases and the technological node gets finer (latest GPUs processed at 5nm), we can estimate that the environmental impacts of graphics cards production increases. 

We study the evolution of the characteristics of graphics cards over time to test this hypothesis. We focus on the leading provider for workstation graphics cards, NVIDIA. We curate a dataset of the 174 workstation graphics cards models released between 2013 and 2024 based on the TechPowerUp GPU database,\footnote{TechPowerUp \url{https://www.techpowerup.com/gpu-specs/} (last accessed 3/17/25) granted authorization to use and share graphic cards data as part of our research project.
}
a Wikipedia page listing NVIDIA graphics cards\footnote{\url{https://en.wikipedia.org/wiki/List_of_Nvidia_graphics_processing_unit} - accessed 3/6/25; content available as CC BY-SA 4.0.} and NVIDIA's published datasheets to settle source disagreements. 

The TechPowerUp database allowed to collect data on 173 cards. The wikipedia page allowed to collect data on 84 cards. 
We merged the two data sets to cross-validate the specifications of the cards. This validated information on 83 out of the 173 models ($47\%$ of the models) in the TechPowerUp data set. One model (Tesla P6) was included in our Wikipedia data set and not in our TechPowerUp data set, and has been added to the final data set.
In cases of divergent information, NVIDIA's published datasheets are taken as reference. These datasheets have also been used to validate information on the compute power of the most popular graphics cards (the P100, V100, A100 and H100 card families). Table~\ref{tab:distribution_workstation} details the temporal distribution of the release dates of the cards included in the dataset. 

The main information gathered for each model includes: Release date, die area, technological node, memory type, memory size, \textit{Thermal Design Power} (TDP) and compute power (Single, Double, Tensor and Half floating precision). Card weight and dimensions were generally not documented in the data sources we consulted and are thus not included in the dataset. The final dataset will be available as the article is released. 

\paragraph{non-NVIDIA-Workstation Graphics Cards.}

We also gather information on a selection of other graphics card models historically used for training. We relied on different sources: the Epoch AI data set on machine learning training hardware \citep{EpochMachineLearningHardware2024} provided details for NVIDIA non-workstation cards (11 card models), the Google Cloud Platform documentation and publications by Google on their hardware, and website of the manufacturers, press releases and benchmarks for the other cards (Cerebras CS-2, Huawei Ascend 910 and AMD Instinct MI250X). Most recent TPUs could not be included because Google does not discole enough information about the technical characteristics of these equipments.
Our final data set is available with the accompanying code.


\subsection{Environmental assessment of graphics card production}

In this paper we follow a cradle-to-gate attributional \gls{LCA} approach~\citep{Hauschild2018life} to assess the environmental impact of graphics cards production. \gls{LCA} seeks to assess impact over the life cycle of a product or service, including production, usage and end-of-life, using a wide variety of environmental impact categories, including but not limited to carbon footprint. 
The goal of an attributional \gls{LCA} is to devise a comprehensive assessment of the environmental burden of the product or service. 

\paragraph{Aims and Scope}

A \emph{Functional Unit} (FU)
is used to define the need fulfilled by the systems under study. 
We perform cradle-to-gate assessments meaning that we account for all phases up to the manufacturing of the card, including raw material extraction and processing, upstream transportation and manufacturing. 
We use the following FU: \textit{producing one graphics card}.

We perform all environmental assessments using attributional \gls{LCA}, adapting the parametric model implemented in MLCA~\citep{Morand2024MLCA}.
The tool aims to cover a representative set of environmental impact categories. For \gls{ICT} hardware and services, carbon footprint, metallic resource depletion, water consumption and toxicity to human and non-human life are all important sources of impact. 
MLCA provides carbon footprint and metallic resource depletion assessments, based on open-source \gls{LCA} information from~\citet{Boavizta2021serveur}.\footnote{Recent open quality information on toxicity and water consumption of ICT equipment and data center facilities is missing~\citep{li2025makingaithirstyuncovering,Lei2022wue}. End-of-life information is also lacking ~\citep{Ficher2025eol}.}

Carbon footprint is assessed according to \emph{Global Warming Potential} (GWP),\footnote{We use the IPCC standard GWP100, corresponding to impact at a time horizon of 100 years.} measured  in kg\COtwo{} for the emissions of greenhouse gases such as carbon dioxide, methane and nitrous oxide~\citep{IPCC2023earth}. 
Metallic resource depletion is assessed through \emph{Abiotic Depletion Potential for elements} (ADPe) which is obtained by assessing the quantity of metal 
(e.g., copper, gold and rare earths) used to produce hardware. 
An aggregated indicator is created based on the amount of each metal and metal rarity compared to Antimony, the standard reference.
The final indicator is expressed in \emph{kilograms antimony equivalent} (kg\Sbe{}) \citep{CML2002}.

\paragraph{Parametric Modeling}

\begin{figure}[ht]
    \centering
    \includegraphics[width=\linewidth]{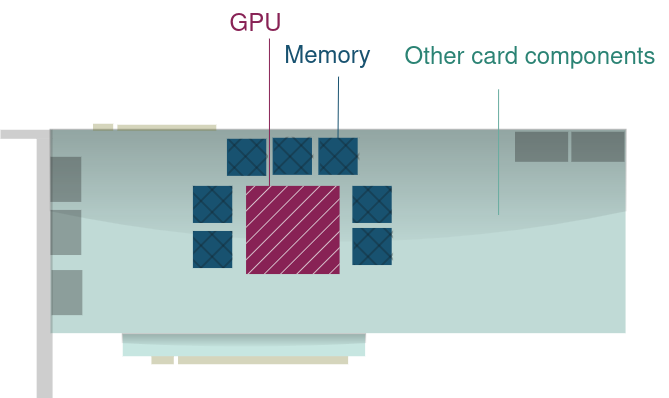}
    \caption{Graphics card modeling.}
    \label{fig:schema_modelisation_MLCA}
\end{figure}

\begin{table*}[ht]
    \centering
    \begin{tabular}{crcp{.4\linewidth}}\toprule
                Parameter & Value & Source & Detail\\\midrule
                $\text{gpu die}_{GWP_{per-cm^2}}$ & $1.97$ kg\COtwo{} & \multirow{2}{*}{\citep{Morand2024MLCA}} & \multirow{2}{\linewidth}{\small Adapted from \citep{Groger2021green}, using data on 14 nm Intel CPU production in Ireland in 2016-2017} \\
                $\text{gpu die}_{ADPe_{per-cm^2}}$ & $5.8 \times 10^{-7}$ kg\Sbe{} & & \\
                memory density & $1.87625$ GB/cm$^2$ & \citep[Table 66]{Groger2021green} & {\small average value for 2015-2021 for 20-18nm node} \\
                $\text{memory die}_{GWP_{per-cm^2}}$ & $3.45$ kg\COtwo{} & \multirow{2}{*}{\citep[Table 42]{ADEME2026GPU}} & 
                \\
                $\text{memory die}_{ADPe_{per-cm^2}}$ & $9.51 \times 10^{-8}$ kg\Sbe{}& &  \\
                $\text{base}_{GWP}$  & $33.381$ kg\COtwo{} & \multirow{2}{*}{\citep[Table 19]{ADEME2026GPU}} & \multirow{2}{\linewidth}{\small adapted from A100 SXM4 40GB} \\
                $\text{base}_{ADPe}$ & $6.05 \times 10^{-3}$ kg\Sbe{} &  & \\\bottomrule
    \end{tabular}
    \caption{Detail of the parameters used in our graphics cards production impact modeling}
    \label{tab:parameters}
\end{table*}

The parametric model in MLCA is based on a bottom-up modeling of hardware to estimate production (and also possibly usage) impacts of hardware. Bottom-up modeling means that hardware is modeled as the sum of its parts.
Figure~\ref{fig:schema_modelisation_MLCA} represents the modeling of hardware in MLCA. Graphics cards 
are modeled based on the size of the GPU (in hatched purple in the Figure), the size of on-board memory (in cross-hatched blue) and constant impact for other card components (in green).
This modeling is translated in the impacts evaluation by adding together the impacts computed for the GPU die, the impacts computed for the memory, and a base impacts to account for all the other components as follows\footnote{where impact $\in$ \{GWP, ADPe\}.}:

\begin{align*}
    \text{graphics card}_{impact} & = GPU_{die_{size}} \times \text{gpu die}_{impact_{per-cm^2}} \\
                                  & + memory_{size} \times memory_{impact_{per GB}}   \\
                                  & + base_{impact}
\end{align*}

The base constant impact accounts for the card casing, heat sink, \textit{Printed Circuit Board} (PCB) and components upstream transportation. Memory impacts are assessed by estimating the memory die area needed the memory volume using a memory density factor\footnote{This factor is constant in our model for lack of data on the evolution of memory \gls{IC} density, even if it is to be expected that density would increase over time along with miniaturization.} (in GB / cm$^2$) and then multiplied by impacts of producing a centimeter square of memory die: 

\begin{align*}
    memory_{impact_{per GB}} & = \frac{\text{memory die}_{impact_{per-cm^2}}}{\text{memory density}}
\end{align*}

Table~\ref{tab:parameters} details the data used in our parametric model. In the assessments, $GPU_{die_{size}}$ and $memory_{size}$ are retrieved from our dataset on graphics cards.


\subsection{Software Used in the Analysis}
Data manipulation and statistical analyses have been performed using emacs Org mode 9.1.9, Python 3.12.3 using the pandas 2.1.4+dfsg and beautifulsoup4 4.12.3 libraries, and R version 4.3.3 (2024-02-29) with the stringr\_1.5.1, tidyr\_1.3.1, lmtest\_0.9-40, zoo\_1.8-14, ggplot2\_3.5.2 and dplyr\_1.1.4 libraries.  Platform: x86\_64-pc-linux-gnu (64-bit) Running under: Ubuntu 24.04.4 LTS

Consistent with open science practices, details on the sources, processes and methodological choices will be made available with the article 

\section{Evolution of Hardware Production Impact}
\label{sec:impact-shifting}


\subsection{Evolution of Graphics Cards Characteristics}

\begin{figure}[t]
    \centering
    \includegraphics[width=0.9\columnwidth]{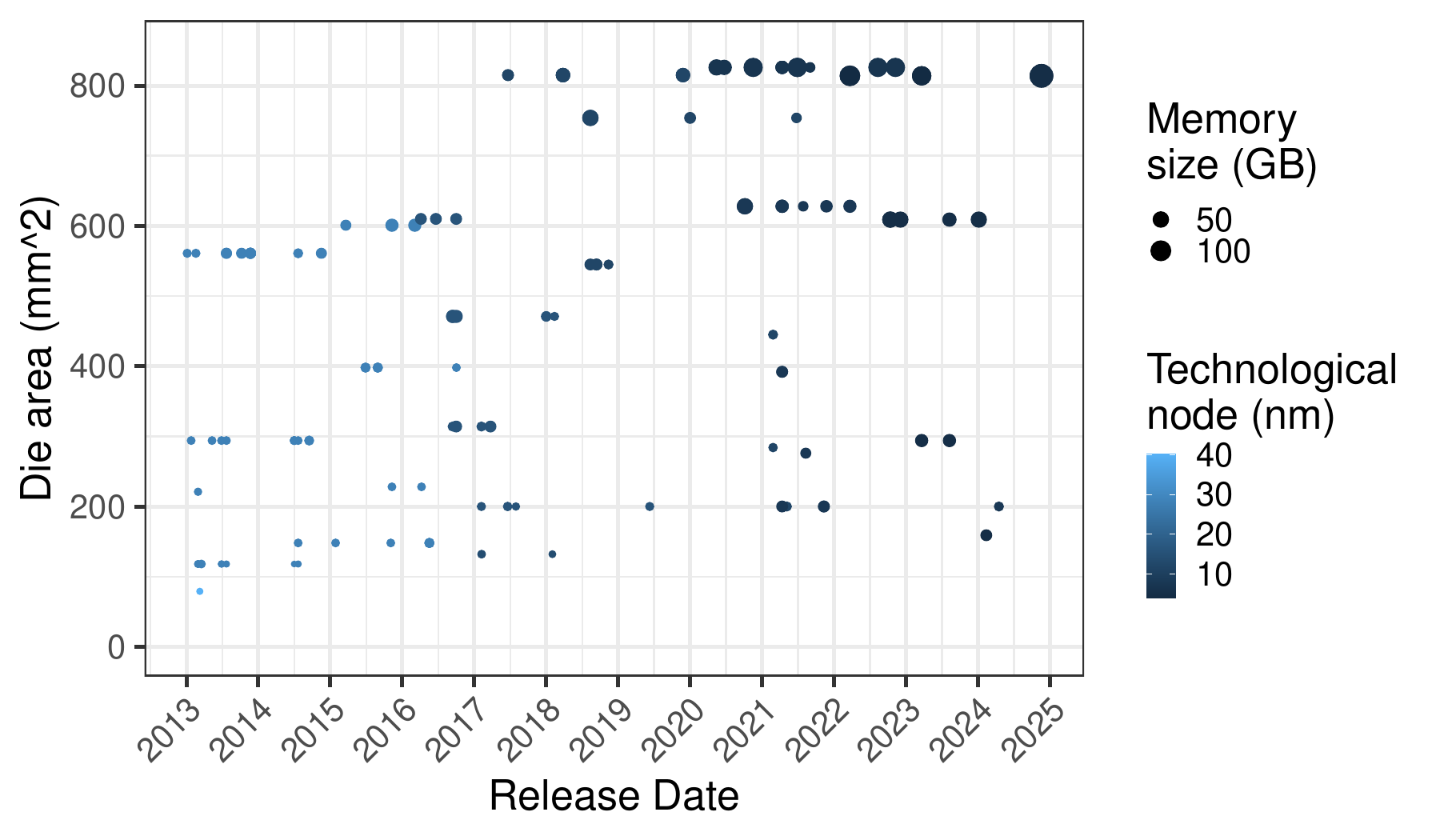}
    \caption{Evolution of the characteristics of NVIDIA workstation graphics cards from 2013 to 2025. Dot size represents memory size and color represents GPU technological node.}
    \label{fig:evol_spec}
\end{figure}

Figure~\ref{fig:evol_spec} shows the evolution of the characteristics of NVIDIA workstation graphics cards from 2013 to 2025. 
Environmental impacts of graphics card production majorly come from the GPU and memory (respectively in hatched purple and in cross-hatched blue in Figure~\ref{fig:schema_modelisation_MLCA}).  The larger the \gls{IC} and the smaller the technological node, the greater the impact. 
Information on the technological node and on die area is available for GPUs while only raw size is available for memory. 
Technological node size has decreased over time and that average die area has increased linearly. 
Memory size (in GB) has grown exponentially (around 30\% Compound Annual Growth Rate). Exponential growth in memory size does not 
lead to 
a proportional increase in memory \gls{IC} area as \gls{IC} have been miniaturized at a pace following Moore's Law.

\begin{figure}[t]
    \centering
    \includegraphics[width=0.9\columnwidth]{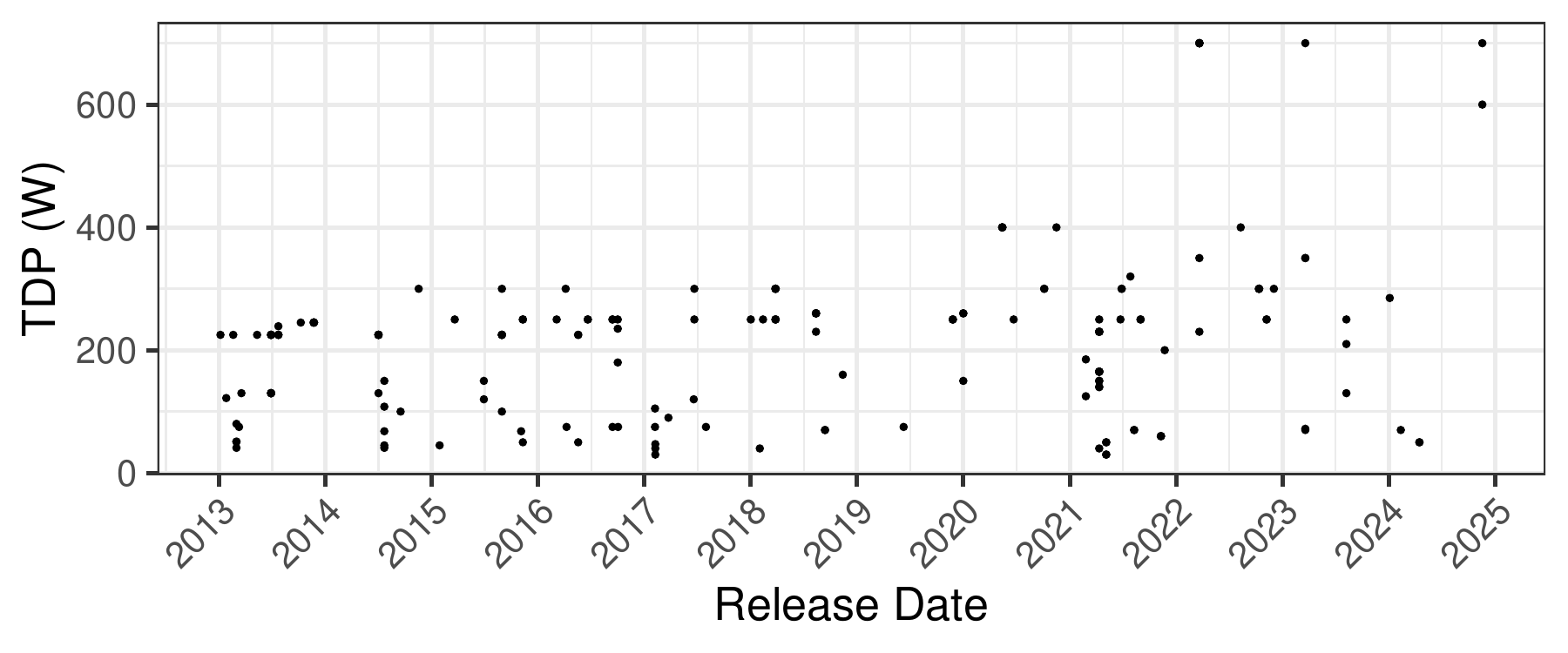}
    \caption{Evolution of the energy consumption of NVIDIA workstation graphics cards (2013-2025).}
    \label{fig:evol_TDP}
\end{figure}

Figure~\ref{fig:evol_TDP} shows the evolution of the energy consumption of NVIDIA workstation graphics cards from 2013 to 2025, in terms of \gls{TDP}. Even if the compute efficiency of the cards has increased exponentially, the total energy consumption of a card has slightly increased over time. This observation 
is consistent with rebound effect, where the energy efficiency improvements on the cards have allowed to increase the number of operations performed on a card at a fixed energy consumption.

\subsection{Graphics Cards Production Damages are Increasing}

Consistent with the observations on graphics cards characteristics, the production carbon footprint of NVIDIA workstation graphics cards increases over time, as presented in Figure~\ref{fig:evol_gpu_GWP}. Metallic resource depletion appears to mostly remains constant over time (Figure~\ref{fig:evol_gpu_ADP}).

\begin{figure*}[ht]
\begin{subfigure}{.45\linewidth}
    \centering
    \includegraphics[width=0.9\columnwidth]{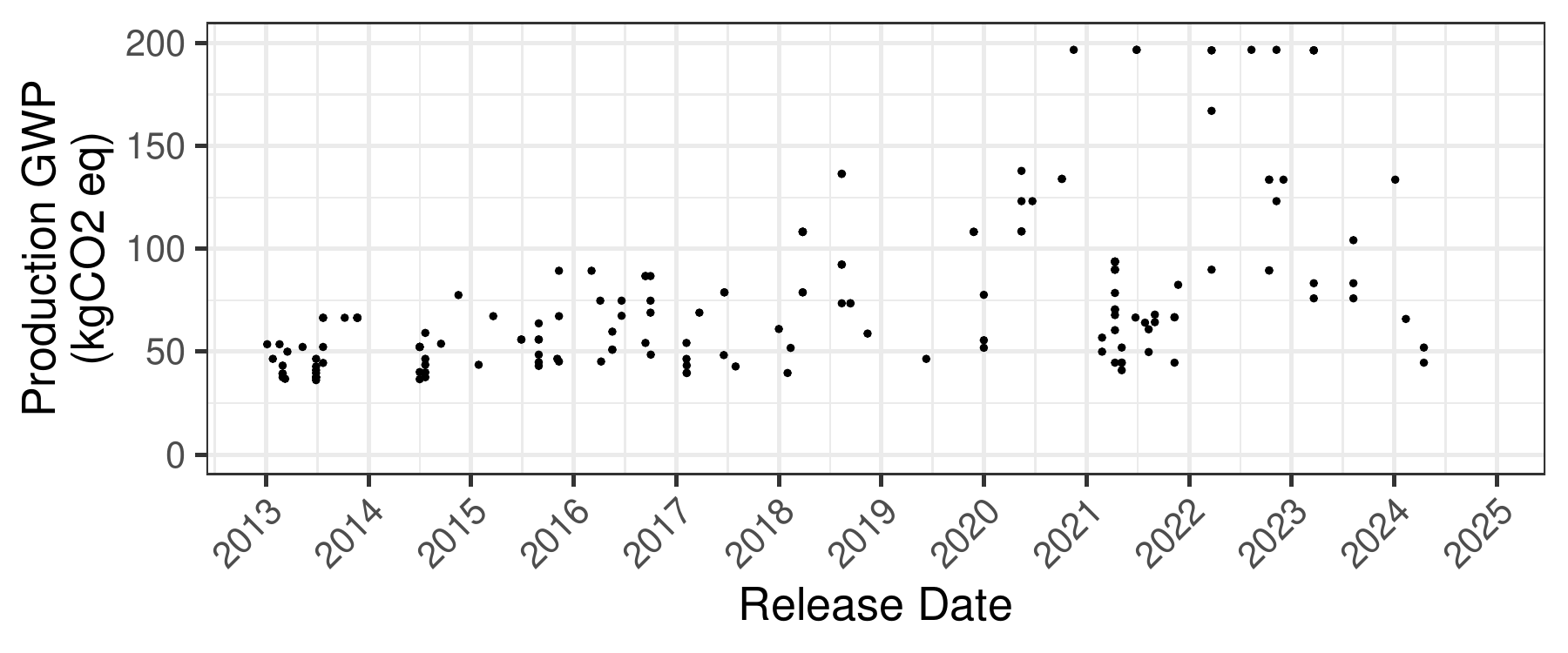}
    \subcaption{GWP}
    \label{fig:evol_gpu_GWP}
\end{subfigure}
\hfill
\begin{subfigure}{.45\linewidth}
    \centering
    \includegraphics[width=0.9\columnwidth]{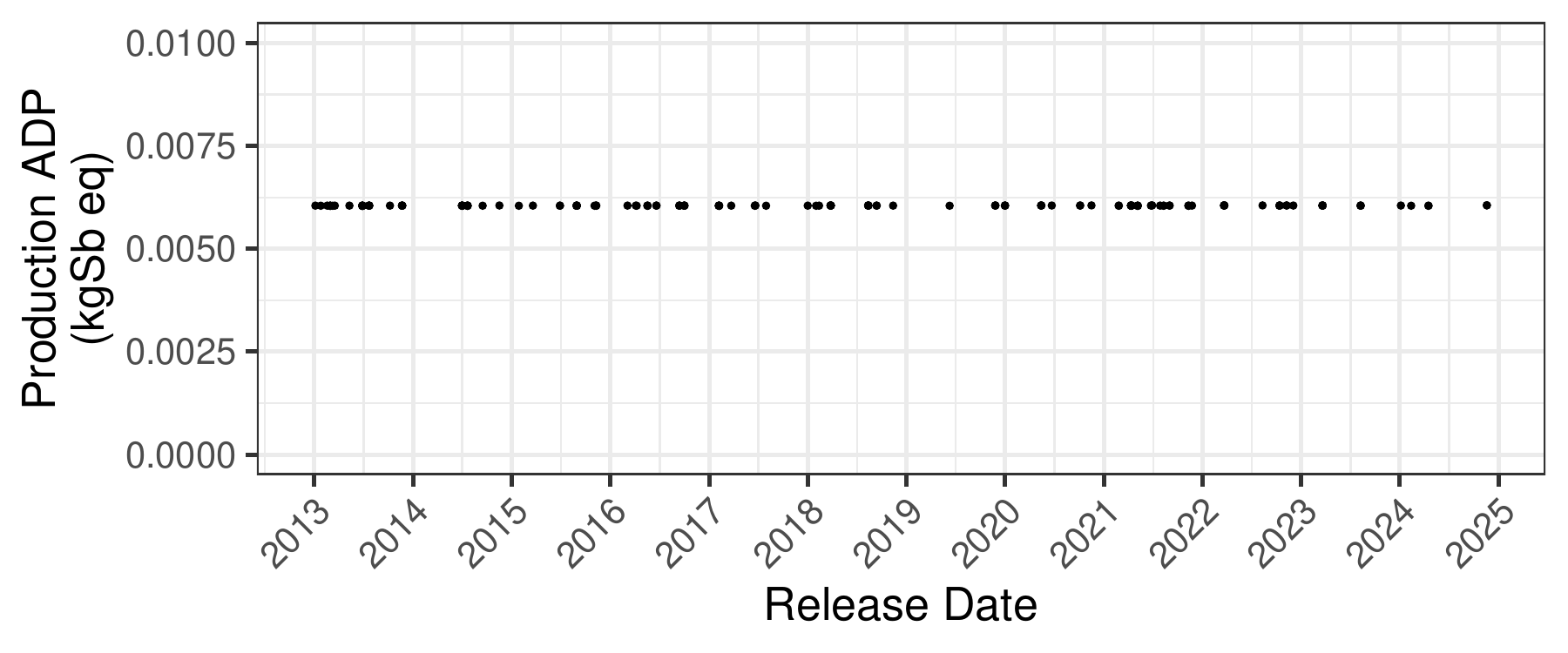}
    \subcaption{ADPe}
    \label{fig:evol_gpu_ADP}
\end{subfigure}
    \caption{Evolution of the production impacts of NVIDIA workstation graphics cards (2013-2025).}
     \label{fig:evol_gpu}
\end{figure*}


\begin{figure*}[ht]
\begin{subfigure}{.45\linewidth}
    \centering
    \includegraphics[width=0.9\columnwidth]{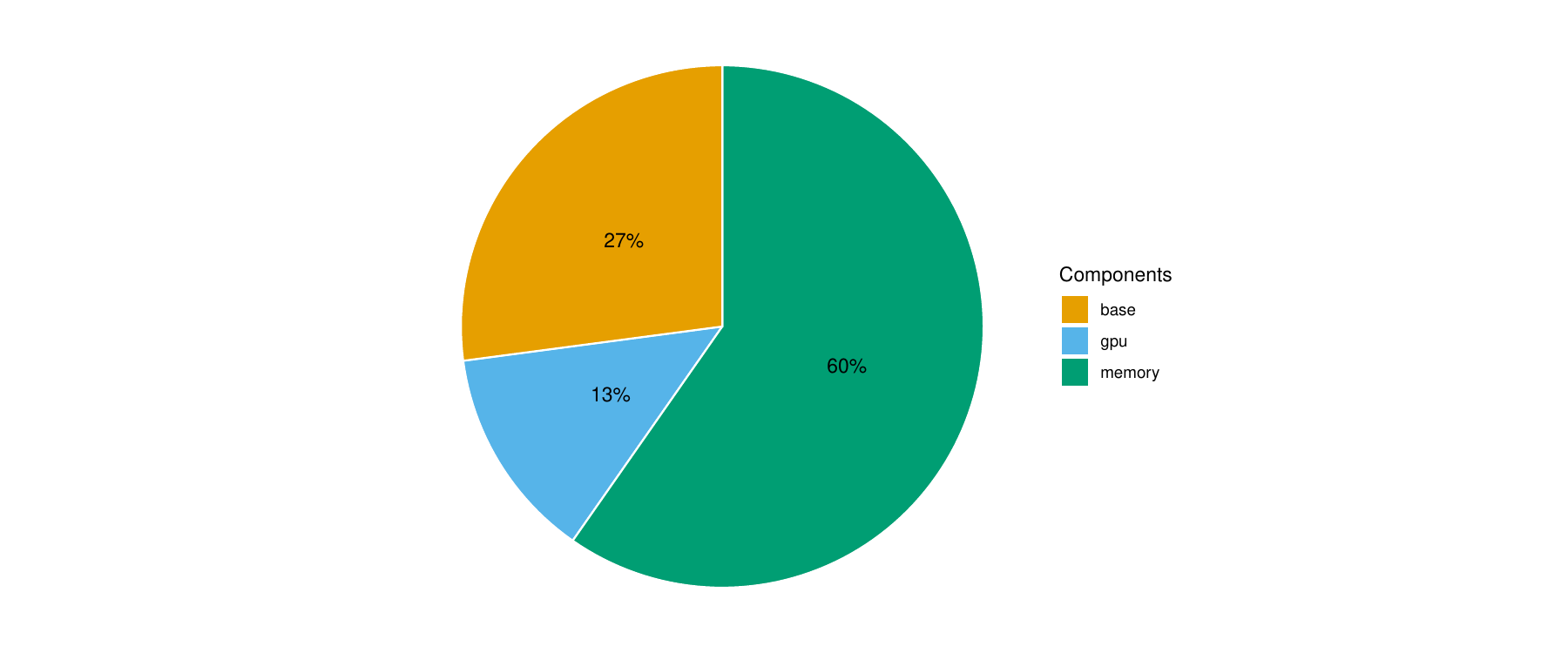}
    \subcaption{GWP}
    \label{fig:evol_gpu_GWP}
\end{subfigure}
\hfill
\begin{subfigure}{.45\linewidth}
    \centering
    \includegraphics[width=0.9\columnwidth]{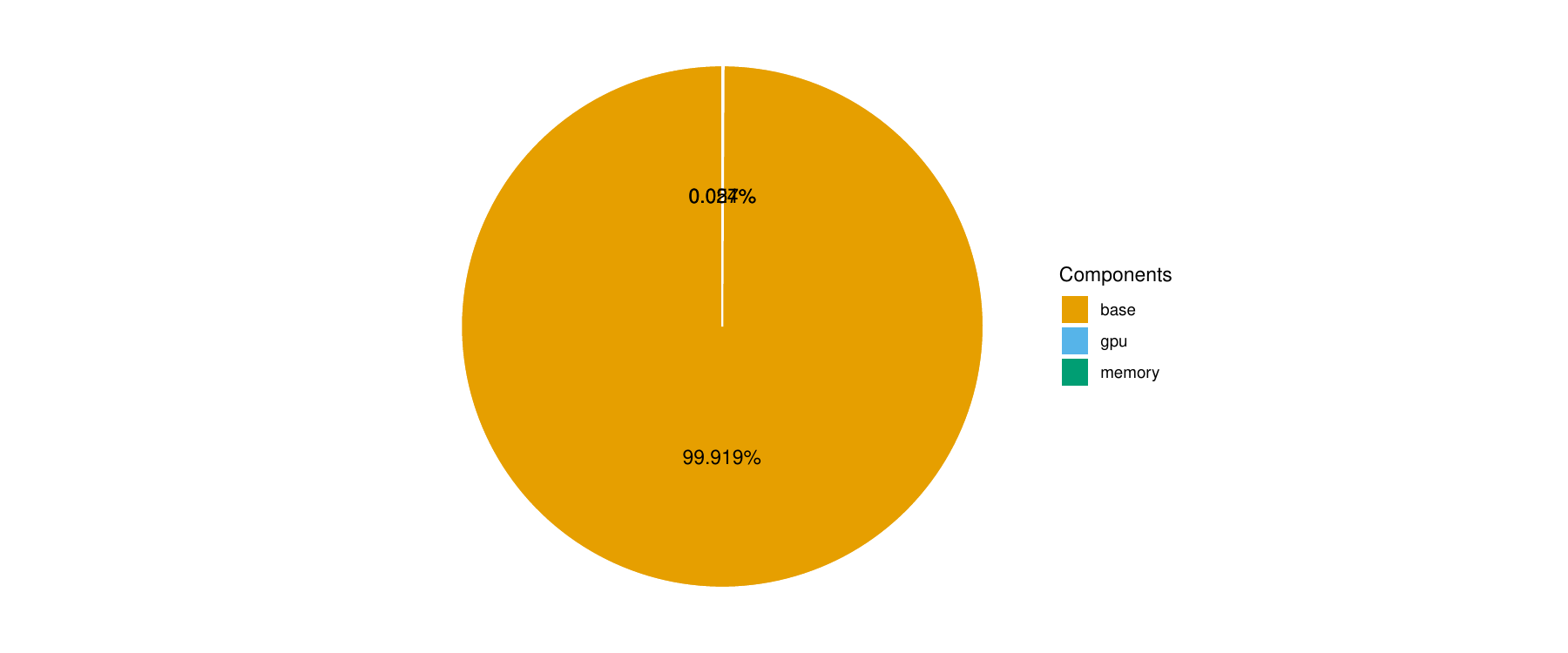}
    \subcaption{ADPe}
    \label{fig:evol_gpu_ADP}
\end{subfigure}
    \caption{Distribution of impact by component for the A100 SXM4 40 GB graphics card}
     \label{fig:share_A100}
\end{figure*}

\KILL{
\begin{table*}[ht]
    \centering
    \begin{tabular}{cccccccc}\toprule
             & &  Minimum   &   First quartile  &     Median    &     Mean  &    Third quartile    &     Maximum \\\midrule
\multirow{3}{*}{\rotatebox[origin=c]{90}{GWP}}  & gpu & 4.30 & 10.6 & 15.4 & 15.0 & 18.9 & 31.8\\
 & memory & 0.76 & 11.2 & 24.1 & 27.7 & 38.4 & 77.0 \\
 & base & 15.5 & 43.1 & 57.1 & 57.3 & 72.8 & 92.7\\\addlinespace

\multirow{3}{*}{\rotatebox[origin=c]{90}{ADPe}} & gpu & 0.005 & 0.019 & 0.037 & 0.032 & 0.043 & 0.074\\
 & memory & 0.10 & 1.57 & 4.58 & 7.04 & 8.75 & 36.0\\
 & base & 63.9 & 91.2 & 95.4 & 93.0 & 98.4 & 99.9\\
 \bottomrule
    \end{tabular}
    \caption{Distribution of shares of impact per component (percentages) for NVIDIA workstation graphics cards}
    \label{tab:distribution_workstation}
\end{table*}
}

\begin{figure*}[ht]
\begin{subfigure}{.45\linewidth}
    \centering
    \includegraphics[width=0.9\columnwidth]{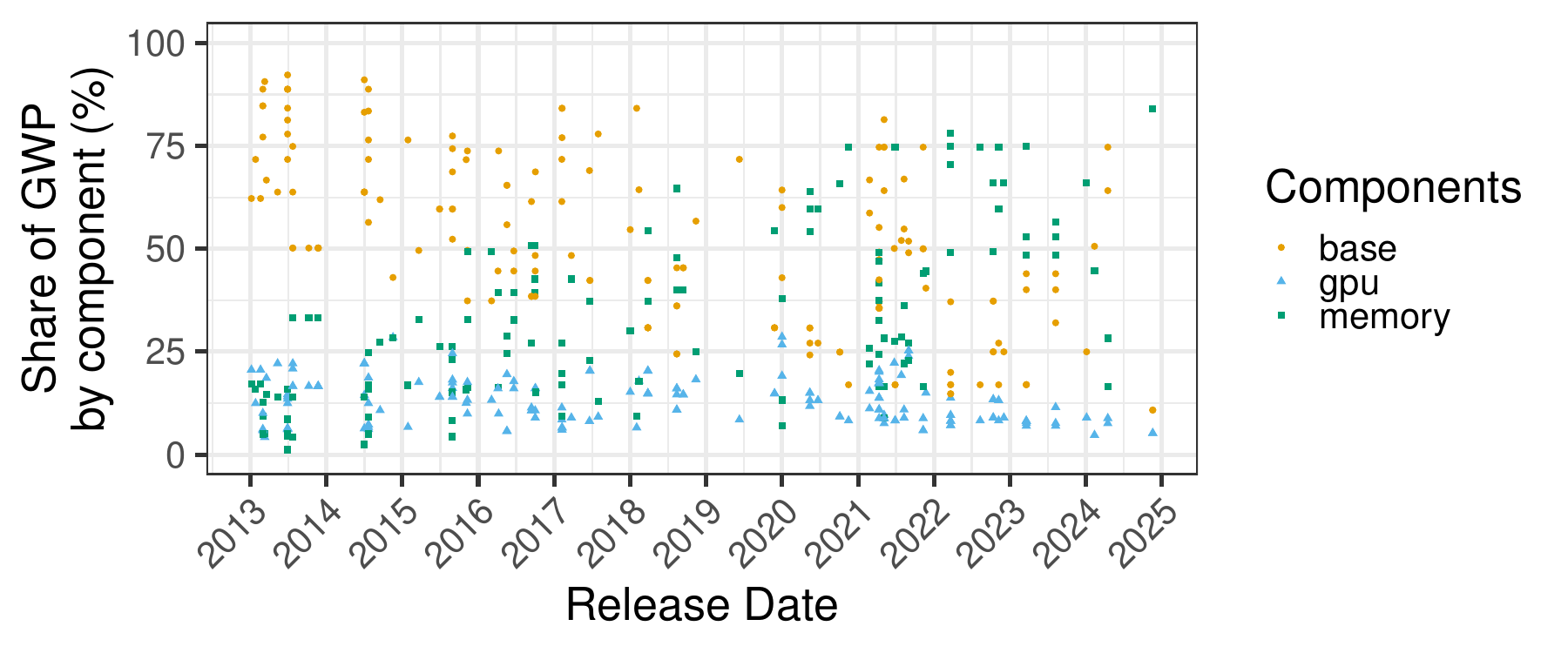}
    \subcaption{GWP}
    \label{fig:evol_gpu_GWP}
\end{subfigure}
\hfill
\begin{subfigure}{.45\linewidth}
    \centering
    \includegraphics[width=0.9\columnwidth]{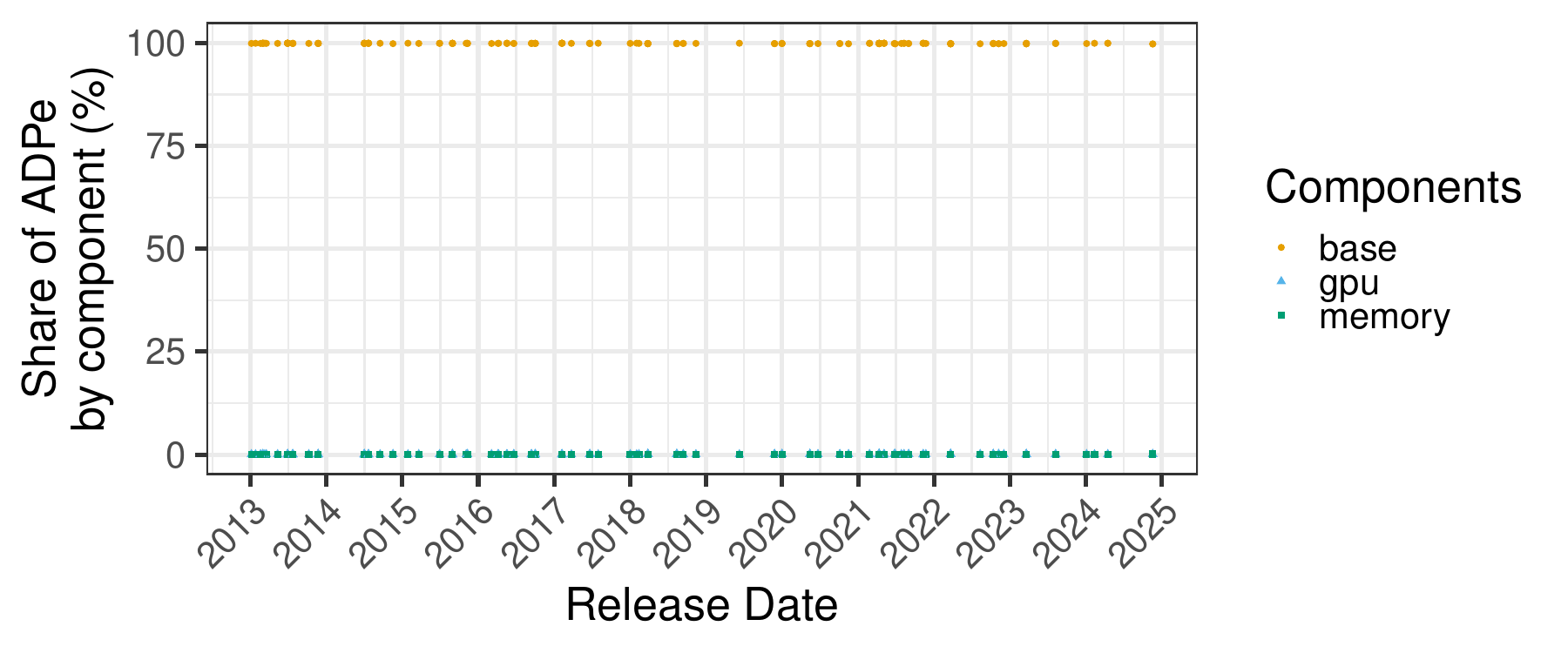}
    \subcaption{ADPe}
    \label{fig:evol_gpu_ADP}
\end{subfigure}
    \caption{Evolution of the assessed share of production impacts by components for NVIDIA workstation graphics cards (2013-2025).}
     \label{fig:evolution_share_impacts}
\end{figure*}

Figure~\ref{fig:evolution_share_impacts} represents the evolution in the assessed shares of impacts for NVIDIA workstation graphics cards over the last decade. 
Base impacts represent the quasi totality of metallic resource depletion for all cards.
Both GPU and memory impacts are increasing over time, leading to a reduction in the share of base impacts. Memory size is the fastest growing factor leading to an increase in the share of impacts from memory and a relative stagnation of the share of GPU impacts over time. Memory becomes the dominant source of impacts in the last years for GWP while base impact remain dominant for ADPe, as illustrated for the case of the A100 SXM4 40 GB detailed in Figure~\ref{fig:share_A100}. GPU represent a significant share of impacts for GWP, but are dwarfed by the other components for ADPe. 

\subsection{Assessments Appear Consistent with Previous Studies}

\begin{table*}[ht]
    \centering
    \begin{tabular}{cccccccc}\toprule
                \multirow{3}{*}{Card model} & \multirow{3}{*}{Release Year} & \multicolumn{2}{c}{Our study} & \multicolumn{2}{c}{ADEME study~\citep{ADEME2026GPU}} & \multicolumn{2}{c}{Difference}\\\cmidrule(lr){3-4}\cmidrule(lr){5-6}\cmidrule(lr){7-8}
                & & GWP & ADPe & GWP & ADPe & GWP & ADPe\\
                & & (kg\COtwo{}) & (kg\Sbe{}) & (kg\COtwo{}) & (kg\Sbe{}) & (\%) & (\%) \\\midrule
L4 & 2023 & 83.3 &  0.006050 & 57.5 & 0.00248 & 31.0 & 59.0 \\
RTX A4500 & 2021 & 82.5 & 0.006052 & 99.6 & 0.00431 & -20.7 & 28.8 \\
A100 SXM4 40 GB & 2020 & 123 &  0.006054  & 126.0 & 0.00605 & -2.27 &  0.06 \\
Titan RTX & 2018 & 92.4 &  0.006053 & 123.0 & 0.00567 & -33.2 & 6.32 \\
GEFORCE GTX 1080 Ti & 2017 & 62.9 &  0.006050 & 73.1 & 0.00522 & -16.2 & 13.7 \\
Tesla P100 SXM2 & 2016 & 74.8 & 0.006051 & 65.3 & 0.00465 & 12.7 & 23.2\\\bottomrule
    \end{tabular}
    \caption{Comparison of environmental assessments for graphics card production between this and ADEME studies}
    \label{tab:comparison_ADEME}
\end{table*}

Table~\ref{tab:comparison_ADEME} presents a comparison of the results obtained in this study versus the ones in for 6 different card models~\citet{ADEME2026GPU}. 
Our parametric model overall produces consistent assessment with the ADEME data. Carbon footprint assessment are mostly centered around the ADEME one, with differences less than a third. Metallic resource depletion tends to be overestimated by our model, but orders of magnitude are consistent and differences are mostly below 50\% with a maximum difference of 7.5\%.

Distribution of the shares of impact seem mostly consistent with GPU and memory dominating carbon footprint, and base impacts representing close to the totality of metallic resource depletion. If we take a closer look at the A100 SXM4 40GB card, the information reported in~\citet{falk2025carboncradletograveenvironmentalimpacts} show a roughly 90 - 10 \% distribution of carbon footprint between GPU and memory at the production stage, and heat sink and PCB impact representing the quasi totality of metallic resource depletion. These results are consistent with the distribution of impact obtained using our model as detailed in Figure~\ref{fig:share_A100}, even if base impacts represent a slightly higher portion of carbon footprint than in ~\citep{falk2025carboncradletograveenvironmentalimpacts}.

\citet{NVIDIA2025pcfH100} provide some information on the production damages of the NVIDIA HGX H100, a baseboard including 8 H100 cards. Their analysis includes transport contrary to our model, but it accounts for 0.4\% of the total, so this inclusion should not biais the comparison. Dividing their assessed total of 1,312 kg\COtwo{} for the GWP of the baseboard gives a value of 164 kg\COtwo{} for one card (including a share of the motherboard). No data is provided on other environmental damages such as resource depletion, so we cannot compare ADP results. Our model assesses the carbon footprint of one H100 card production at 225 kg\COtwo{}. The two values are of the same magnitude, even if the NVIDIA value is a little bit more conservative. 

Table~\ref{tab:comparison_NVIDIA} details the comparison on the distribution of damages over the different components between our model and NVIDIA's data. Most of the difference can be explained by memory, which is estimated at a higher value than in NVIDIA's study, contrary to the gpu and base which are estimated lower than in NVIDIA's study. Still, the results between the two study are of the same magnitude for the different components and the relative shares between components are consistent. One possible explanation for the observed difference would be the use of different LCA database between this and NVIDIA study.  

\begin{table*}[ht]
    \centering
    \begin{tabular}{cccccc}\toprule
                \multirow{3}{*}{Component} & \multicolumn{2}{c}{Our study} & \multicolumn{2}{c}{NVIDIA study \citep{NVIDIA2025pcfH100}} & Difference\\
                & GWP & share & GWP & share & GWP\\
                & (kg\COtwo{}) & (\%) & (kg\COtwo{}) & (\%) & (\%) \\\midrule
memory & 176 & 78.1 & 68.25 & 41.8 & 61.2\\
gpu & 16 & 7.1 & 41,5 & 25.4 & -159 \\
base & 33.4 & 14.8 & 53.3 & 32.7 & -59.5\\
total & 225 & 100 & 164 & 100 & 27.1\\
\bottomrule
    \end{tabular}
    \caption{Comparison of environmental assessments for the carbon footprint of one H100 graphics card production between this and NVIDIA studies}
    \label{tab:comparison_NVIDIA}
\end{table*}



\subsection{Hardware Procurement are Rapidly Increasing}

\begin{figure}[ht]
    \centering
    \includegraphics[width=0.9\columnwidth]{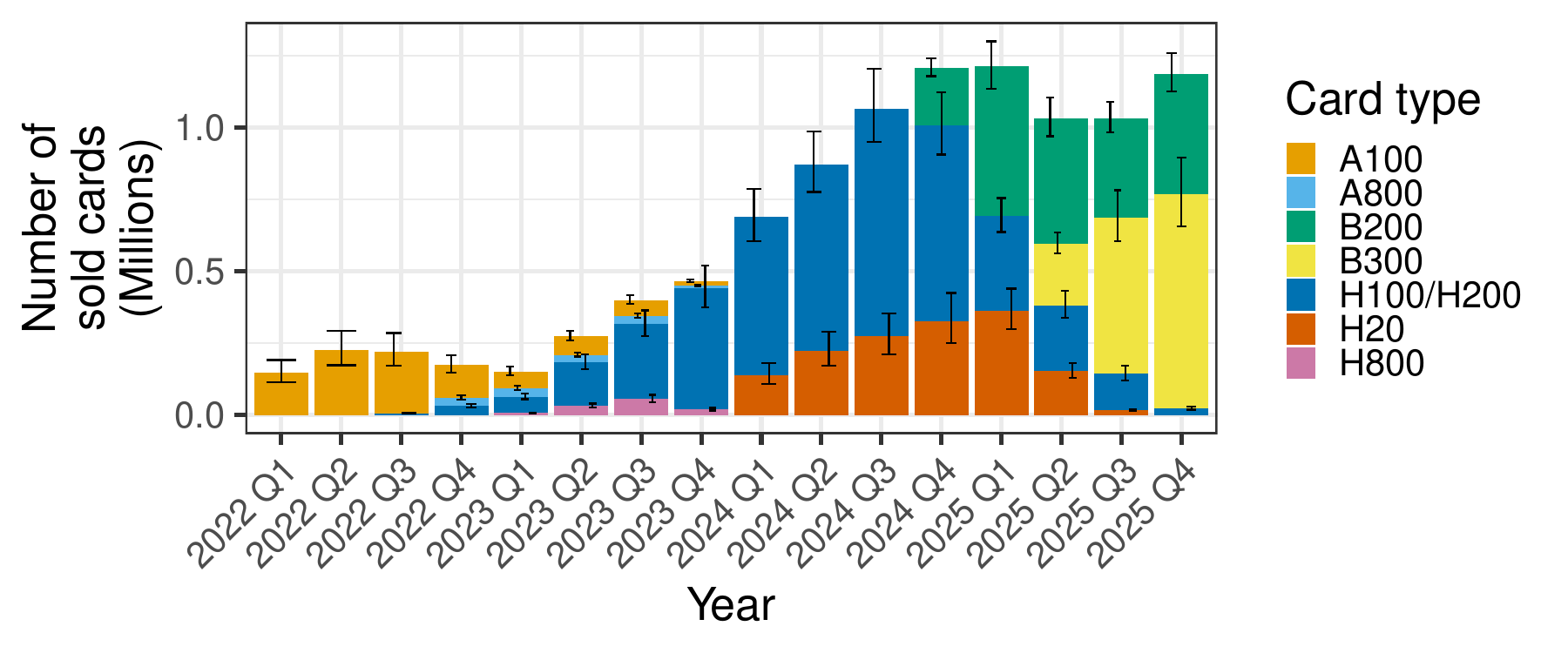}
    \caption{Evolution of the number of NVIDIA AI cards sold in each quarter (2022-2025), adapted from \citet{EpochAIChipSales2026} data}
    \label{fig:evolution_nb_card}
\end{figure}

We present insights on the global environmental implications of graphics card procurement. To that end, we mobilize the data from  \citet{EpochAIChipSales2026} on graphics cards sales from 2022 to 2025. We combine this data with our production damage assessments to obtain insights on the evolution of graphics cards production damages at a global scale. Figure~\ref{fig:evolution_nb_card} shows the evolution in the number of AI cards sold by NVIDIA during each quarter from 2022 to 2025. It shows a large increase in card procurement from around 200,000 units sold in Q1 2022 to more than a million at the end of 2024. Sales appear mostly stable at a high level of more than a million unit sold during each quarter of 2025. We can also see A100 initially dominating sales, being replaced by H100/H200 in 2023 and themselves being replaced by B200 and B300 cards starting at the end of 2024.

\begin{figure*}[ht]
\begin{subfigure}{.45\linewidth}
    \centering
    \includegraphics[width=0.9\columnwidth]{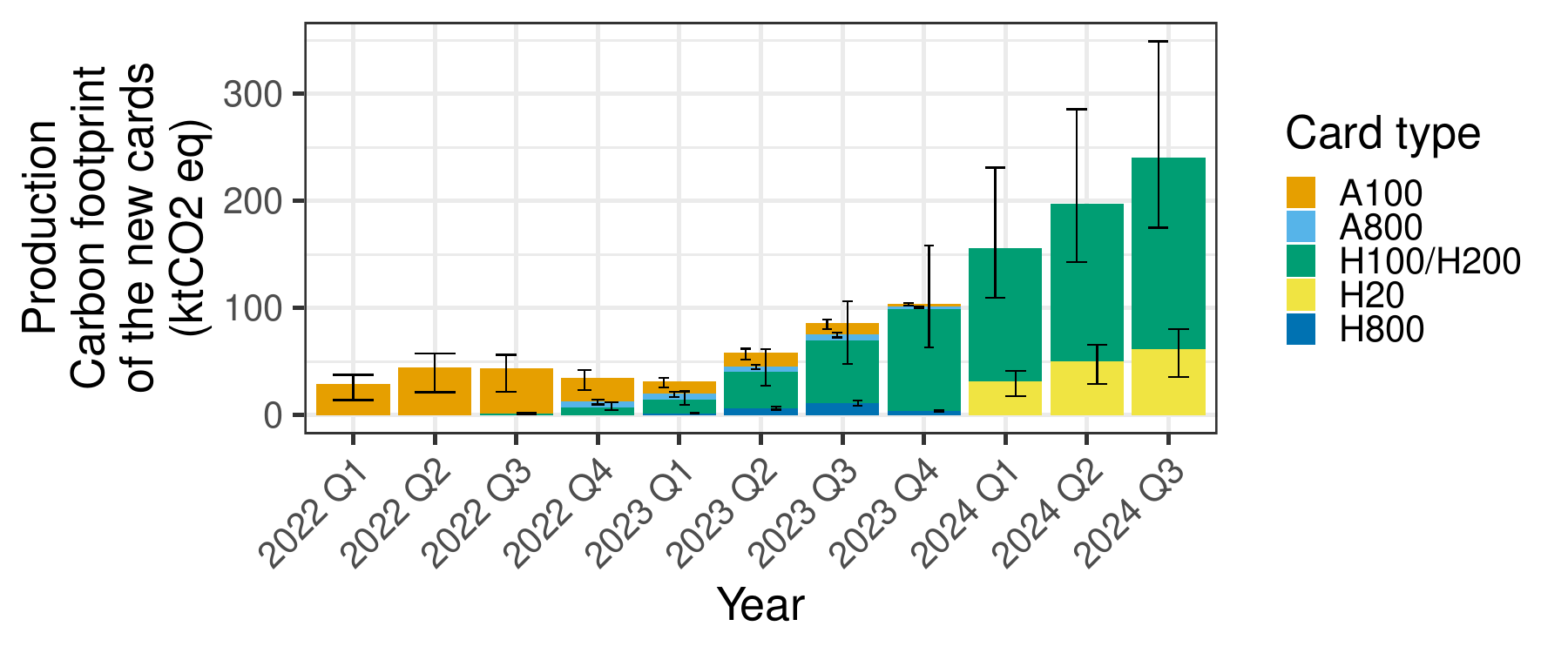}
    \subcaption{GWP}
    \label{fig:evol_new_gpu_GWP}
\end{subfigure}
\hfill
\begin{subfigure}{.45\linewidth}
    \centering
    \includegraphics[width=0.9\columnwidth]{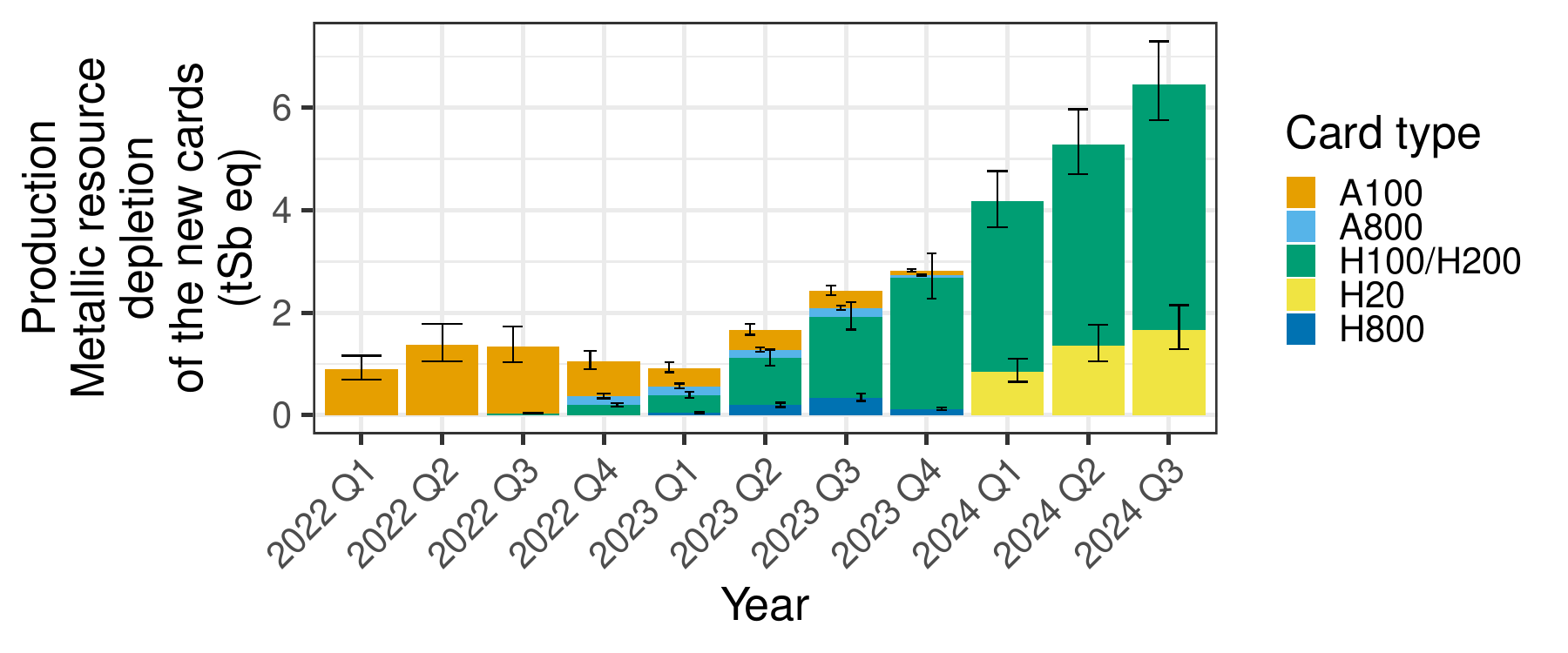}
    \subcaption{ADPe}
    \label{fig:evol_new_gpu_ADP}
\end{subfigure}
    \caption{Evolution of cumulative production damages of NVIDIA AI graphics cards (2022-2024). Error bars represent uncertainties around the number of sales, as well as variations between models of the same generation (e.g., A100 SXM4 40 GB VS. A100 SXM4 80 GB). t stands for the metric ton.}
     \label{fig:evol_new_gpu_damages}
\end{figure*}

As our data set does not comprise information on the most recent B200 and B300 cards, due to lack of transparency notably on the die areas for these new models, we now focus on the period from Q1 2022 to Q3 2024. Figure~\ref{fig:evol_new_gpu_damages} shows the evolution of environmental damages from the production of the sold NVIDIA AI graphics cards over the period. Consistent with the stark increase in the number of sales, and the increase in damages per card it shows a large increase in damages from card production at a global scale both for carbon footprint and for metallic resource depletion. Damages should continue increasing even if sales volume remain stable from 2025 onward since newer generations of cards have higher unitary production damages.

We now compare the total production footprint of the card sold in a year (Q4 2023 - Q3 2024) with planetary boundaries, as in \citet{Morand2024MLCA} using the data from \citet{Serenella2020environmental}. This comparison is intended to better grasp the magnitude of the assessed values in the context of global sustainability limits. The total estimated GWP is 698 (min 430, max 1,074) kt\COtwo{}. This amounts to the maximum emissions of 708 (min 437, max 1,091) thousand people in a scenario respecting the planetary boundary for GWP. The total estimated ADPe is 18.7 (min 15.6, max 22.5) t\Sbe{}. This amounts to the maximum metallic resource extraction of 591 (min 492, max 711) thousand people in a scenario respecting the planetary boundary for ADPe. The magnitude of 700 thousand people corresponds to the municipal population of a city such as København (Copenhagen), Denmark \citep{enwiki:1354926906}. 


\section{Discussion}

\subsection{High Environmental damages of Graphics Cards}


Our study shows increasing environmental production damages from the production of individual graphics cards over the last decade, with a carbon footprint growing from around 50kg\COtwo{} to around 200kg\COtwo{} per unit for recent models, and metallic resource depletion assessed at around 0.01 kg\Sbe{} per unit. 
This makes producing a single workstation card as impacting as the production of a complete laptop or even desktop computer~\citep{Gupta2020chasing}. Even if (AI) chips for user devices are likely to have smaller capacities and thus smaller production footprint than their workstation counterparts, their inclusion in user devices probably significantly increases the production footprint of these devices. Compute servers often comprise up to 8 graphics cards, making their production as impacting as the rest of the whole server~\citep{Morand2024MLCA}. 

Furthermore, AI graphics cards sales have been starkly rising over the last few years, leading to a large increase in the environmental damages of AI graphics card production at a global scale, even if only accounting for NVIDIA chips. 
NVIDIA sold AI graphics card production in 2024 was assessed to represent the maximal pollution of the population of a city large as København (Copenhagen), Denmark, in a scenario respecting planetary boundaries. While this value may appear rather small compared with the whole ICT sector, AI graphics cards are a supplementary burden to the ICT sector environmental damages and are not substituting to other devices. Furthermore, AI applications are yet to prove that they can effectively serve climate change mitigation on a large scale~\citep{Falk2024Challenging,Ligozat2021unraveling,Kaack2021aligning}. The globalization of AI also appears to mostly promote incremental techno-fixes approaches to sustainability while "hampering confrontational tactics and grassroots resistance, especially in authoritarian states"~\citep{Dauvergne2021globalization}.
Graphics card production and its current sales growth is thus in itself a significant unsustainability contributor. It ought to be questioned to address the pressing environmental crises such as climate change and resource depletion.

The bulk of life-cycle damages incurred by AI graphics cards is located in their use phase, especially for greenhouse gas emission, due to their high energy consumption and intense usage patterns~\citep{falk2025carboncradletograveenvironmentalimpacts}. Still, mitigation strategies focused on electricity consumption will not suffice to address the unsustainability of AI, as production is an already very significant.
Changing usage electricity consumption does not mitigate hardware production damages, which are also increasing.
Worse, The geographic changes intended to reduce carbon intensity could incentivize
shorter lifespan for data center facilities~\citep{Velkovka2015impermanent},  leading to higher damages of these facilities.

\subsection{Study Scope Supports Reliable Trends}

Our parametric model has limitations for production damage assessment, including not accounting for technological node and assuming fixed memory density. Still, estimates appear consistent with recent research~\citep{ADEME2026GPU,falk2025carboncradletograveenvironmentalimpacts}. Results for recent cards also are on par with the ones for Google TPUs~\citep{Google2025acvTPU}.

Our parametric model uses easy to access information on all graphics cards.
The use of constant impact factors for the GPU die, memory density and base impact means that our model will tend to under-estimate the impacts for the most recent models, while over-estimating impacts for older ones. 
Improving the precision of our model would require accessing sufficient data to account for further variations in graphics cards composition and production processes. 
Base impacts could for instance vary depending on the card dimension and weight, as proposed in the ADEME model~\citep{ADEME2026GPU}. These information are however more complex to access and are more often than not not documented in the data sources we used to construct our dataset. Base impact could also benefit from being differentiated by type (e.g., PCIe vs. SXM4), that are documented in our dataset and present differences like for connectors. We however lack data to account for that factor.
Our dataset documents the technological node used for producing the different GPUs, access to quality open data on the impacts associated with producing dies at different nodes would be needed to account for these variations. 
Accounting for changes in memory die size and production would require detailed data on the memory density for memory \gls{IC} of different memory types (e.g., GDDR5 vs. HMB2e), which could then be used in combination with the documented memory types in our dataset.
Still, our model provides mostly consistent estimates with existing studies on graphics card production impacts, and allows for easy assessments for other cards.

Most recent cards like NVIDIA B200 and B300 might be challenging to account for since they embed both GPUs and CPUs. Upgrading the parametric model for these new types of cards could be realised by using existing information and models of CPU impacts, for instance as are currently implemented in MLCA and BoaviztAPI~\citep{Morand2024MLCA, Simon2024boaviztapi}.  They are also not included in our dataset because we could not find information notably on the die areas for these cards. It will be possible to include them once manufacturers release more information about these cards. 

\subsection{Damages Go Beyond Carbon Footprint}
\label{subsec:discussion-metrics}

This study examined the carbon footprint and metallic resource depletion of producing graphics cards, finding that both metrics have increased over time. Importantly, AI causes important social and environmental damages beyond these metrics over its whole life cycles~\citep{Tubaro2025dual,Valdivia2025supply,Valdivia2026follow,Falk2024attributional}.
One can for instance think of water usage~\citep{Mytton2021datacenter}, ecosystem destruction~\citep{Comber2023Computing}, and pollution from hardware mining and disposal. AI poses significant social consequences and ethical challenges~\citep{Bender2021parrots,Jiang2023Art}, highlighting the need for a more comprehensive assessment that incorporates qualitative analyses and a broader range of impacts.

\section{Conclusion} 
\label{sec:conclusion}

Graphics cards are a key component of AI infrastructures, which are frequently updated to cope with increasing computation demand and modern software requirements. In this paper, we show that the environmental damages from graphics cards production have increased in the past decade, worsening the damages from these updates. This increase has been worsened by the surge in cards sales.  

Impact reduction thus cannot rely on improving the efficiency of operational damages of AI, and must address the expanding production impacts of hardware.

To meaningfully reduce the environmental damages of AI, we must move beyond efficiency gains and fundamentally reassess the role of AI in a sustainable society, prioritizing environmental limits over computational growth.

\section*{Acknowledgments}

This work has received funding from the French "Agence Nationale pour la Recherche" under grant agreement InExtenso - ANR-23-IAS1-0004. 
Clément Morand was supported by a doctoral grant from ENS Rennes.  
The authors thank TechPowerUp for allowing us to us their data in this study. 
Clément Morand would like to thank Aina Rasoldier for his help in gathering the data from TechPowerUp

\section*{CRediT author statement}

Clément Morand: Conceptualization, Data curation, Methodology, Software, Writing - Original Draft Aurélie Névéol: Conceptualization, Writing- Review \& Editing, Supervision, Funding acquisition Anne-Laure Ligozat: Conceptualization, Writing- Review \& Editing, Supervision, Methodology.

\bibliographystyle{elsarticle-harv} 
\bibliography{biblio.bib}

\end{document}